\documentclass[12pt]{article}
\usepackage{graphicx,psfrag,epsfig}
\usepackage{amssymb,amsmath,amscd,amsthm}
\usepackage{graphicx,psfrag,epsfig}

\usepackage{graphicx}
\usepackage[active]{srcltx}

\newtheorem{theorem}{Theorem}

\newtheorem{lemma}[theorem]{Lemma}

\date{}

\begin{document}

\date{}
\title{Resonance regimes of scattering by small bodies
with impedance boundary conditions}
\author{E. Lakshtanov\thanks{Department of
Mathematics, Aveiro University, Aveiro 3810, Portugal, lakshtanov@rambler.ru}, B.
Vainberg\footnote{Department of Mathematics and Statistics, University
of North Carolina, Charlotte, NC 28223, USA, brvainbe@uncc.edu}}
\maketitle

\begin{abstract}
The paper concerns scattering of plane waves by a bounded obstacle
with complex valued impedance boundary conditions.
We study the spectrum of the Neumann-to-Dirichlet
operator for small wave numbers and long wave asymptotic
behavior of the solutions of the scattering problem.
The study includes the case when $k=0$ is an eigenvalue or a resonance.
The transformation from the impedance to the Dirichlet boundary
condition as impedance grows is described. A relation between
poles and zeroes of the scattering matrix in the non-self adjoint case is established.
The results are applied to a problem of scattering by an obstacle with a springy
coating. The paper describes the dependence of the impedance on the properties of
the material, that is on forces due to the deviation of the boundary of the obstacle
from the equilibrium position.
\end{abstract}


\hspace{0.5cm}


\begin{center}

\textbf{Introduction}
\end{center}

We consider the scattering of plane waves by a bounded obstacle
$\mathcal O \in \mathbb R^3$ with  a smooth boundary $\partial
\mathcal O \in C^2$ and impedance boundary conditions. The
scattered field $u=u(r),~r=(x,y,z)$ satisfies the Helmholtz
equation in $\Omega = \mathbb R^3 \backslash \mathcal O$ and
radiation conditions:
\begin{eqnarray} \label{1}
\left\{\begin{array}{lll} \Delta u(r)+k^2u(r)=0,~~r \in \Omega,~~k > 0,\\
\int_{|r|=R} \left |\frac{\partial u(r)}{\partial |r|}-ik u(r)
\right  |dS=o(1),~~R \rightarrow \infty. \end{array}\right.
\end{eqnarray}
The Robin boundary condition holds at the boundary:
\begin{equation}\label{2}
\frac{\partial u}{\partial n}- \gamma (k) u=-\left (\frac{\partial
e^{ik(r\cdot \alpha)}}{\partial n}-\gamma (k)e^{ik(r\cdot \alpha)}
\right  ),~~r \in \partial \Omega,
\end{equation}
where $\alpha \in S^2$ is the direction of the incident plane wave, $n$ is the exterior normal for $\mathcal O$ (directed into $\Omega$).

In the mathematical literature, the impedance is usually assumed to be purely imaginary with negative imaginary part (see, e.g.,
\cite{coltonKress, Isakov, uhlmann}). In \cite{AlRamm}, \cite{smallRamm}
one can find the theorem on the existence of the solution to the problem
(\ref{1})-(\ref{2}) with $\Im \gamma \leq 0$ and the long wave
asymptotics away from resonance regimes. We are not imposing any
restrictions on the real or imaginary parts of $\gamma$. Recall
that $\gamma$ is determined by the forces at the boundary of the obstacle.
Existence of an elastic component of the force (proportional to
the displacement) defines the real part of $\gamma$. A friction
(the force proportional to the velocity of the displacement)
defines the imaginary part of $\gamma$. A more detailed analysis
of the dependence of ã on the forces is given in Appendix I. Thus, all the values of
the impedance are of interest, including those which correspond to
artificial forces such as "negative friction" or negative
compressibility coefficient.

For the sake of transparency, we assume that the impedance does not depend on the point of the boundary. We also will assume that $\gamma = \gamma(k)$ is analytic in $k$ in a neighborhood of the point $k=0$. The latter preserves the analyticity of the scattering matrix and will allow us to discuss poles. Using perturbation arguments one can easily extend the main results below to the case of sufficiently smooth $ \gamma(k)$.

Let us recall that any solution of the problem (\ref{1}) has the following asymptotic behavior at infinity:
\begin{equation}\label{12}
u(r)=u_\infty \left(\theta\right )\frac{e^{ik|r|}}{|r|} + o\left
(\frac {1}{|r|} \right ),~\theta=\frac{r}{|r|},~ \quad |r|
\rightarrow \infty.
\end{equation}
Function $u_\infty \in L_2(S^2)$ is called the scattering amplitude (it depends also on $k$ and $\alpha$), and the square of its norm
\begin{equation} \label{crs}
\sigma_k(u)=\int_{S^2}
|u_\infty(\theta)|^2dS,
\end{equation}
is called the total cross-section.

Problem (\ref{1})-(\ref{2}) can be easily reformulated in terms of the Neumann-to Dirichlet operator
$\mathcal D= \mathcal D(k)$ which maps the normal derivative $\frac{\partial u}{\partial n }|_{\partial
\Omega}$ of the Neumann problem for the equations (\ref{1}) into the value $u |_{\partial\Omega}$ of
the solution at the boundary. Let
$u(r)\in H^2_{loc}(\Omega)$ satisfy (\ref{1}). Then $\frac{\partial u}{\partial n }|_{\partial
\Omega}
\in H^{1/2}(\partial \Omega),~u |_{\partial\Omega}\in H^{3/2}(\partial \Omega)$ and $\mathcal D(k)$ is
a bounded operator from $H^{1/2}(\partial \Omega)$ into $H^{3/2}(\partial \Omega)$. In fact, it is a
pseudo-differential operator of order $-1$ (see, e.g., \cite{vainberg}) and will be considered as a compact
operator in $L_2(\partial \Omega)$:
\begin{equation} \label{operD}
\mathcal D(k):L_2(\partial \Omega)\rightarrow L_2(\partial
\Omega),~~k \geq 0.
\end{equation}
If operator $\mathcal D(k)$ is considered for complex $k \in \mathbb C$ it is understood as an analytic continuation of (\ref{operD}).

From the Green formula it follows that
\begin{equation}\label{dirGreen}
u(r)= -\int_{\partial \Omega} G_k(r,s) \frac{\partial u(s)}{
\partial n} dS(s), \quad r,s \in \partial \Omega,
\end{equation}
where $G_k(r,s)$ is the Green function of the Neumann problem (see \cite{ts} for the existence of $G_k$).
Thus, problem  (\ref{1})-(\ref{2}) for $u\in H^2_{loc}(\Omega)$ is equivalent to the equation
 \begin{equation}\label{spectr}
  \left (I -\gamma(k) \mathcal D \right )v=- \left . \left
(\frac{\partial e^{ik(r \cdot \alpha)}}{\partial n} - \gamma(k)
e^{ik(r \cdot \alpha)} \right ) \right |_{\partial \Omega}, \quad
v \in H^{1/2}(\partial \Omega),
  \end{equation}
and the corresponding field is given by (\ref{dirGreen}) with $\frac{\partial u}{\partial n }=v$.

Let $\{\varsigma_j(k),~j=0,1,2,...\}$ be the set of eigenvalues of
operator $\mathcal D(k)$, and let $\{\varphi_j(k)\}$ be the
corresponding eigenfunctions. Usually we will not mark the
dependence of these functions $\varphi_j(k)=\varphi_j(k,r)$ on
$r\in\partial\Omega$.

We study the spectrum of the operator $\mathcal D(k)$, solvability and
properties of the solution of the scattering problem (\ref{1})-(\ref{2}) and of the total cross-section $\sigma_k(u)$. In particular,
it is shown that the eigenvalues $\{\varsigma_j(k)\}$ of the operator
$\mathcal D(k)$ are analytic in a neighborhood of $k=0$, and the
inverse values $\{\varsigma_j^{-1}(k)\}, k \geq 0,$ belong to a
half strip in the upper complex half plane $\{z: 0< \Im z < c(k), \Re z  < d(k) \}$. Long
wave asymptotic of the solution of (\ref{1})-(\ref{2}) is
obtained and the location of resonances is described. We show
convergence of the impedance problem to the Dirichlet problem when the
impedance grows along any ray through the origin different from $(0,-\infty)$. These general results are
applied to a scattering problem for an obstacle coating by a
springy layer. Some other results related to topics discussed below can be found in \cite{lc, arx}.

The paper is organized as follows. The main results are stated in the next section. The proofs are given in section 3 followed by an attachment containing additional discussions concerning springy covers.
\begin{center}
\textbf{Main Results.}
\end{center}
The following theorem is rather simple, but very essential. A proof will be given in the next section.
\begin{theorem}\label{td}
1. Condition $\frac{1}{\gamma(k)} \not \in Sp(\mathcal D(k))$ implies the existence and uniqueness of the
solution of problem (\ref{1})-(\ref{2}).

Operator $\mathcal D(k)$ is meromorphic in $k \in  \mathbb C$ with all the poles located in the lower half plane.
 Let $\gamma(k)$ be analytic in some connected domain $U\subset \mathbb C$, and $\frac{1}{\gamma(k')}
\not \in Sp(\mathcal D(k'))$ for some value of $k=k'\in U$. Then the solution of problem (\ref{1})-(\ref{2}) exists and is meromorphic in $k \in U$ as an element of $H^2_{loc}(\Omega)$.

2. The eigenvalues $\{\varsigma_j(k)\}$ and eigenfunctions
$\{\varphi_j(k),~j=0,1,2,...\}$ of $\mathcal D(k)$ are analytic in
some neighborhoods of the origin $k=0$. To be more exact, for any
$\varepsilon>0$ there exists a neighborhood $U_{\varepsilon}$ of
the origin in the complex $k$-plane such that  $L_2(\partial
\Omega)$ can be represented as a sum
\begin{equation}\label{split}
L_2(\partial \Omega)=L_{1,\varepsilon}(k)+L_{2,\varepsilon}(k),~ k\in U_{\varepsilon},
\end{equation}
where the projection operators $P_i(k):L_2(\partial \Omega)\rightarrow L_{i,\varepsilon}(k)$
are analytic in $k\in U_{\varepsilon}$ and orthogonal when $k=0$, and the first space  is finite
dimensional and has the form:
 \[
 L_{1,\varepsilon}(k)=\text
 {span} \{\varphi _j(k),~ 0 \leq j \leq N_{\varepsilon}\}.
 \]
Here,  eigenfunctions $\varphi _j(k)$ are analytic in $k \in U_{\varepsilon}$ (together with their
eigenvalues $\varsigma_j(k)$), linearly independent for each $k$ and orthogonal when $k=0$.
The second space is also invariant for $\mathcal D(k)$, and the norm of the restriction of
$\mathcal D(k)$ on $L_{2,\varepsilon}(k)$ does not exceed $\varepsilon:~||\mathcal
D(k)|_{L_{2,\varepsilon}(k)}||<\varepsilon,~k\in U_{\varepsilon}$.

3. For each $k >0$, the set of inverse values
$\{\varsigma_j^{-1}(k)\}$ belongs to a half strip in the upper complex
half plane $\{z:0<\Im z<c(k), ~\Re z < d(k)\}$, with the only limiting point at
$-\infty$.
\end{theorem}
The last statement can be found in \cite[4.2]{agr}.
We will show that the last statement of the theorem leads to the
convergence of the solutions of the impedance problem
(\ref{1})-(\ref{2}) to the solution of the corresponding Dirichlet
problem if the impedance grows along any ray through the origin different from $(0,-\infty)$. Namely, the following theorem holds.
\begin{theorem} \label{teorD}
Let $u_t$ be the solution of (\ref{1})-(\ref{2}) with $\gamma
=te^{i\delta}$, where $-\pi<\delta <\pi]$, and
let $w$ be the scattered field in the case of the Dirichlet
boundary condition, i.e. $w$ satisfies (\ref{1}) and
$w=-e^{ik(r\cdot \alpha)}$ on $\partial\Omega$. Then
 \begin{equation}\label{perehod}
\lim_{t \rightarrow \infty} \|u^t-w\|_{L_2(\partial \Omega)}=0,
\quad \lim_{t \rightarrow \infty} \|u^t_\infty-w_\infty
\|_{L_2(S^2)}=0.
  \end{equation}
\end{theorem}
An analog
of Theorem \ref{teorD} with $k=0$ was proved recently in
\cite{alberrammN}.

Now we are going to discuss the long wave asymptotics ($k
\rightarrow 0$) of the solution of problem (\ref{1})-(\ref{2}).
Since the problem is equivalent to (\ref{spectr}), it is obvious
that the result depends on the asymptotics of functions $1-\gamma
(k)\varsigma_j(k)$ as $k\rightarrow 0,$ where
$\varsigma=\varsigma_j(k),~ j=0,1,2,...,$ are eigenvalues of
operator (\ref{operD}). The impedance $\gamma (k)$ can be an
arbitrary function, and therefore first we will study the asymptotics of the
functions $\varsigma=\varsigma_j(k)$. Then we impose certain
conditions on the difference $1-\gamma (k)\varsigma_j(k)$ and
obtain some results for the solution of (\ref{1})-(\ref{2}).

Note that operator $\mathcal D(0)$ is symmetric, and therefore
numbers $\varsigma=\varsigma_j(0)$ are real (which is not
necessarily true for $\varsigma_j(k),~ k >0 $) . Moreover,
$\varsigma_j(0)<0$ since operator $\mathcal D(0)$ is negative, see
(\ref{Grf0}). We always can enumerate the eigenvalues in such a
way that the sequence $\{\varsigma_j(0)\}$ is not decreasing.
Recall that the eigenfunctions $\{\varphi_j(0)\}$ of the operator
$\mathcal D(0)$ form an orthonormal basis in
$L_2(\partial\Omega)$. They also can be chosen to be real valued.
The eigenfunctions $\{\varphi_j(k)\}$ for complex $k, |k|\ll 1$,
are analytic in $k$ due to Theorem \ref{teorD}. Let $u_0$ be the
unit function on $\partial\Omega$ and let $c_j$ be the scalar
projection of $u_0$ on $\varphi_j$, i.e.,
\begin{equation}\label{a3}
c_j:=\int_{\partial \Omega} \varphi_j(0) dS. \quad j =
0,1,2,\ldots
\end{equation}
Then
\begin{equation}\label{cp0}
\sum_{j=0}^\infty c_j^2=|\partial \Omega|.
\end{equation}
Note that eigenfunctions $\varphi_j(0)$ which correspond to a multiple eigenvalue of operator $\mathcal D(0)$ can not be chosen absolutely arbitrary, since the existence of analytic in $k$ continuation leads to some restriction in their choice.
\begin{theorem} \label{t3}
1. The smallest eigenvalue $\varsigma_0(0)$ is simple and $c_0 \neq
0$. The following estimate holds
\begin{equation}\label{cp}
|\varsigma_0(0)|\leq \frac{S}{4\pi C},
\end{equation}
where $S$ is the area of $\partial\Omega$ and $C$ is its
electrostatic capacity.

2. The following relation holds
\[
\varsigma'_j(0)=\frac{-i}{4\pi}(c_j)^2, \quad j=0,1,2,\ldots
\]

3. Let the impedance $\gamma(k)$ be analytic in a neighborhood of
$k=0$ and  $\frac{1}{\gamma(0)}\notin \{\varsigma_j(0),
j=0,1,\ldots\}$. Then the solution of (\ref{1})-(\ref{2}) exists in
some neighborhood of $k=0$ and is analytic in $k$ as an element of
$H^2_{loc}(\Omega)$.

Let $\gamma(k)$ be analytic in a neighborhood of $k=0$ and
$\frac{1}{\gamma(0)}=\varsigma_p(0)$, where
$\varsigma_p(0)=\varsigma_{p+1}(0)=...=\varsigma_{p+m-1}(0)$ is an
eigenvalue of $\mathcal D(0)$ of multiplicity $m \geq 1$, and
$\varphi_{j},~ p \leq j \leq p+ m-1,$ are the corresponding
eigenfunctions. Let
\begin{equation}\label{gpr}
\gamma'(0) \neq \frac{i(c_{j})^2}{4\pi (\varsigma_j(0))^2}, \quad
~p \leq j \leq p+ m-1.
\end{equation}
 Then the solution of (\ref{1})-(\ref{2}) exists
and depends meromorphically (as an element of $H^2_{loc}(\Omega)$) on $k$ in some neighborhood of $k=0$. It has
a pole of the first order at $k=0$, and the value of $u$ at the
boundary has the form
\begin{equation}\label{ass}
u(r)=\frac{1}{k}\sum_{j=p}^{p+m-1}b_j\varphi_j(0,r)+ O(1), \quad r
\in
\partial \Omega, \quad k \rightarrow 0,
\end{equation}
where
\begin{equation}\label{bjDef}
b_j=\frac{-c_j  }{(
\gamma\varsigma_j)'(0)}=\frac{c_j\varsigma_j(0)}{\frac{i}{4\pi}(
c_j)^2-{\gamma'(0)(\varsigma_j(0))^2 }}.
\end{equation}
The scattering amplitude has the form
\begin{equation}\label{scatamplt}
u_{\infty}(\theta)=\frac{\gamma(0)}{4\pi k}\sum _p^{p+m-1}c_jb_j +
O(1), \quad k \rightarrow 0.
\end{equation}
The total cross-section has the form
\begin{equation}\label{assSigma}
\sigma_k=\frac{\sigma^0}{k^2}+O\left (\frac{1}{k} \right ), \quad
k \rightarrow 0;~~~\sigma^0=\frac{1}{4\pi} \left
|\sum_{j=p}^{p+m-1}\gamma(0)c_jb_j \right |^2 .
\end{equation}

4. For any $\varepsilon>0$ and $U_{\varepsilon}, N_{\varepsilon}$ defined in theorem \ref{td}, there exists a $\gamma$-independent (in the class of impedances analytic in some fixed neighborhood of $k=0$) domain
$V_{\varepsilon} \subset U_{\varepsilon}$ such that the solution $u$ of the scattering problem (\ref{1})-(\ref{2}) and the total cross-section $\sigma_k(u)$
have poles at all points $k \in V_{\varepsilon}$, where $\gamma(k)\varsigma_j(k)=1$ for some $j \leq N_{\varepsilon}$ with $c_j \neq 0$.
\end{theorem}

\textbf{Remarks.}  1. When $k=0$ is a
pole, the main part of the scattering amplitude does not depend on
the angle of scattering  $\theta \in S^2$ (see
(\ref{scatamplt})), so \textit{the scattering is isotropic}.

2. In the case of a sphere, the growth of the total cross section in
resonance regimes is well studied in acoustics (see, e.g.,
\cite{leighton}). It was noted that the scattering of acoustic waves by small
air bubbles in the water has total cross section which is more than 500
times larger than
geometrical cross section of the bubble.

Order of the poles in (\ref{ass}) and (\ref{assSigma}) at $k=0$
was restricted by the assumption (\ref{gpr}). In general, these
poles may have any order which is defined by the order with which
functions $1-\gamma(k)\varsigma_j(k)$ vanish at $k=0$. There is
another condition which restricts the order of these poles.

Recall that the scattering matrix is the operator
\begin{equation}\label{scat}
\mathcal S=I+\frac{ik}{2\pi}F \quad : \quad L_2(S^2) \rightarrow
L_2(S^2),
\end{equation}
where $S^2$ is the unit sphere in $R^3, ~~F$ is the operator with the integral kernel $u_{\infty}=u_{\infty}(\theta,\alpha)$:
\[
(Fh)(\theta)=\int_{S^2}u_{\infty}(\theta,\alpha)h(\alpha)dS.
\]
Operator $\mathcal S$ is unitary when $\gamma$ is real, $k>0$ (see, e.g.,
\cite{grinberg}). Obviously, the latter is equivalent to the
relation
\begin{equation}\label{unitar}
\mathcal SF^*=\left (I+\frac{ik}{2\pi}F \right )F^*=F,~~ \Im
\gamma=0,~ k>0.
\end{equation}
The unitarity of $\mathcal S$ and (\ref{scat}), (\ref{crs}) immediately imply the following statement.
\begin{theorem}\label{unit}
Let $\gamma(k)$ be analytic in a neighborhood of $k=0$ and $\Im\gamma(k)=0, ~k\geq 0.$ Then the scattering amplitude has a pole at $k=0$ of at most the first order and the total cross-section has a pole at $k=0$ of at most the second order.
\end{theorem}
\textbf{Example.} Consider the case when $\mathcal O =B$ is the ball of radius $R=1$ centered at the
origin. Recall that functions
\[
u_{n,m}(kr)=h_n(k|r|)Y_{n,m}(\theta),~~h_n(\rho)=H^{(1)}_{n+1/2}(\rho)/\sqrt{\rho}
 \]
 satisfy (\ref{1}), and their restrictions on the sphere $\partial B$ are eigenfunctions of $\mathcal D(k)$. Here $Y_{n,m},~m\leq n,$ are spherical functions (eigenfunctions of the Laplace-Beltrami operator with the eigenvalues $\mu_n=-n(n+1)$) and $H^{(1)}_{n+1/2}$ are the Hankel functions. The solution of the problem (\ref{1})-(\ref{2}) can be obtained by the Fourier method:
 \[
 u=\sum_{n=0}^{\infty}\sum_{m=-n}^n \frac{f_{m,n}(k)}{kh_n'(k)-\gamma(k)h_n(k)}u_{n,m}(kr),
 \]
 where $f_{n,m}$ are the coefficients of the expansion of the right-hand side $f$ in equation (\ref{spectr}) in the basis $\{Y_{n,m}\}$. The same result will be obtained if problem (\ref{1})-(\ref{2}) is solved by means of (\ref{spectr}), (\ref{dirGreen}).

 It is convenient to enumerate the eigenfunctions $\varphi_{n,m}$ of  $\mathcal D(k)$ by two indexes:
$\varphi_{n,m}=Y_{n,m}/||Y_{n,m}||$, and introduce $c_{n,m}$ instead of $c_p$ (see (\ref{cp})). Then  $c_0=c_{0,0}=\sqrt{4\pi}$ and $c_{n,m}=0$ for $(n,m)\neq(0,0)$. Since $f=1$ when $k=0$, we have
$f_{0,0}=1,~f_{n,m}=0,~(n,m)\neq(0,0)$. Operator  $\mathcal D(k)$ has eigenvalues
\[
\varsigma_n(k)=h_n(k)/kh'_n(k)
\] of multiplicity $2n+1$. In particular, $\varsigma_0(k)=1/(-1+ik) $ and
\[
\varsigma_n(0)=-\frac{1}{n+1}.
\]

The field $u$ may have a pole of any order at $k=0$ if $\gamma(k)$
is not real. For example, a pole of a high order can be obtained
if $\gamma(k)=-1+ik+O(k^N)$. If $\gamma(k)\equiv \varsigma_n(0)$
then the order of the pole is restricted by theorem \ref{unit} and
the result depends on whether $n=0$ or $n>0$. If $\gamma(k)\equiv
-1,$ (i.e., $n=0$), then the field and the scattering amplitude have
poles of the first order, and therefore the total cross-section
$\sigma_k(u)$ has a pole of the second order. If $\gamma(k)\equiv
-(n+1),~n>0,$ the scattering amplitude and the total cross-section
do not have poles at $k=0$. The latter is a consequence of the
fact that $c_{n,m}= 0, ~ (n,m) \neq(0,0).$

\textbf{Scattering matrix.} It is  well known that in the self-adjoint
case (in particular, for impedance boundary conditions with a real
valued impedance) the scattering matrix $\mathcal S,~k >0,$ is
unitary, and the relation  $\mathcal S (k)\mathcal S^*(k)=I,~ k
>0,$ implies that the set of poles and the set of zeroes of
$\mathcal S , k\in \mathbb C,$ are complex conjugate. The next
statement generalizes this fact.

We will use the subindex $\gamma$ to indicate the dependence of the scattering matrix $\mathcal S$ and operator F on the impedance.
\begin{theorem} \label{tlast}
1. Let the impedance $\gamma(k)$ be an entire function (not necessarily real valued on $R$).

Then $\gamma_1(k)=\overline{\gamma(\overline{k})}$ is an entire function, and

1) the following relation replaces (\ref{unitar}):
\[
\mathcal S_{\gamma}(F_{\gamma_1})^*=F_{\gamma},~~k>0,
\]

2) the following two statements are equivalent:

$\mathcal S_{\gamma}(k)$ has a pole at $k=k_0 \in \mathbb C.$

$\mathcal S_{\gamma_1}(k)$ has a non-trivial kernel at $k=\overline{k_0}$.

3) $\mathcal S_{\gamma}(k)$ may have a kernel at real $k=k_0 >0$ only in the case of absorbing impedance, $\Im \gamma<0$.
\end{theorem}

\textbf{Remark.} The existence of a non-trivial kernel of $\mathcal S_{\gamma}(k)$ at a point $k=k_0,~0<k_0\ll 1,$ allows one to concentrate energy at a boundary of a small obstacle using an incident wave for which the scattered wave has zero amplitude,
see \cite{mir2} for a practical implementation of this effect. In fact, let $v$ be the solution of (\ref{1})-(\ref{2}) with $-ik$ in radiation condition replaced by $ik.$
Let $v_{\infty}=v_{\infty}(\theta, \alpha)$ be the scattering amplitude of this solution. It is defined by (\ref{12}) with $v$ instead of $u$ and $-ik$ instead of $k$. Then, for each $\alpha$,
\begin{equation}\label{00}
\mathcal Sv_{\infty}(-\theta, \alpha)=u_{\infty}(\theta, \alpha).
\end{equation}
This relation is valid for real $\gamma$, and therefore it is
valid for complex $\gamma$ due to the analyticity of both sides in
$\gamma$. (To show (\ref{00}) for real $\gamma$ we note that
(\ref{unitar}) and the reciprocity identity for $F$ lead to
$S\overline{u_{\infty}}(-\theta, -\alpha)=u_{\infty}(\theta,
\alpha),~~k>0$, which justifies (\ref{00}) for $k>0$, and therefore
for all $k \in \mathbb C$).

The following is an equivalent form of (\ref{00}). Let $u$ be a solution of the problem
\begin{eqnarray} \label{1234}
\left\{\begin{array}{lll} \Delta u(r)+k^2u(r)=0,~~r \in \Omega,~~k > 0,\\
\frac{\partial u}{\partial n}- \gamma (k) u=0,~~r \in \partial\Omega,\\
u(r)=u_{in} \left(\theta\right )\frac{e^{-ik|r|}}{|r|} -u_{out} \left(\theta\right )\frac{e^{ik|r|}}{|r|} +
o(1/|r|),~\theta=\frac{r}{|r|},~ \quad |r| \rightarrow \infty. \end{array}\right.
\end{eqnarray}
Then
\begin{equation} \label{sscc}
\mathcal Su_{in}(-\theta)=u_{out}(\theta),~~k > 0.
\end{equation}

Let now $\Omega$ be the exterior of the ball of radius $r_0\ll 1$ centered at the origin, and let $\gamma=-ik_0$. Then $u=\frac{e^{-ik|r|}}{|r|}$ satisfies (\ref{1234}), $u_{in}\equiv 1$, $u_{out}\equiv 0$ and the field $u$ and the density of the energy in a small neighborhood of the boundary are much bigger than in any other point if $r_0$ is small enough.

\textbf{An obstacle with a springy coating.} The last statement of this section concerns one applied problem: acoustic scattering by an obstacle coated by a springy
material. It is modeled (see Attachment) by problem (\ref{1})-(\ref{2}) with the impedance
\begin{equation} \label{imz}
\gamma(k)=-Zk^2, ~Z\gg 1.
\end{equation}
The value of $Z$ depends on the relative characteristics of the
cover layer and the exterior medium. For example, $Z
\gg 1$ if there is a
gas
 layer around the obstacle with an elastic exterior membrane (for example, rubber) and radial walls in the layer attached to both the obstacle and the membrane, and the whole construction is submerged into a liquid. The walls are needed in order to achieve a springy character of the layer (to localize the output of a point exterior pressure). The attachment contains calculation of $Z$ for this particular construction.

The last
statement of Theorem \ref{t3} implies

\begin{theorem}\label{apl}
For each $i_0<\infty$ there exists $Z_0$ such that the total
cross-section of problem (\ref{1})-(\ref{2}) with impedance
$\gamma(k)=-Zk^2,~ Z \geq Z_0,$ has poles at points
\[
k_i^{\pm}=\pm\frac{1}{\sqrt{Z|\varsigma_i(0)|}}+
\frac{\varsigma'_i(0)}{2Z(\varsigma_i(0))^2}+O \left
(\frac{1}{Z^{3/2}} \right  ), \quad Z \rightarrow \infty,
\]
for all $i\leq i_0$ with $c_i\neq 0.$
\end{theorem}
Note that the value of $\Im \varsigma'_i(0) < 0$ in the formula above is given in item 2 of
Theorem \ref{t3}. Hence, generally (when the number of non-zero
coefficients $c_i$ is infinite) the number of poles in any
neighborhood of $k=0$ tends to infinity when $Z\rightarrow
\infty.$

Theorem \ref{apl} shows a similarity between standard Helmholtz
resonators and the construction discussed above which we will call
"a gas layer in a liquid". The main feature of Helmholtz resonators
is the presence of poles of the scattering cross-section as close
to the point $k=0$ as we please. The same is true for $k_0^{\pm}$
in the case of a gas layer in a liquid.
Moreover, one can deform the Helmholtz resonator into a springy covered obstacle
by changing simultaneously the shape of the resonator and the impedance
in such a way, that the
scattering cross-sections of all the intermediate problems have poles at the same distance from the origin.

%Recall that the Helmholtz resonator is a cavity which can be
%constructed by the following procedure. Consider a closed compact
%surface $S$. Assuming that an $\varepsilon$-neighborhood of a
%point on the surface $S$ can be stretched, we are pressing this
%neighborhood into the domain bounded by $S$ until we get a
%cavity with a thickness of walls of order $\varepsilon$. Let
%$S_{\varepsilon}$ be the final surface (Helmholtz resonator), and
%let $F_s,~s\in [0,1],$ be a family of homomorphisms of the surface
%$S$ such that $F_0=S,~F_1=S_\varepsilon$. The Neumann condition is
%imposed on the boundary $S_\varepsilon$ of the Helmholtz
%resonator. The total cross-section of the exterior problem for the
%Helmholtz resonator with the Neumann boundary condition on
%$S_\varepsilon$ has two poles at the distance of order
%$O(\varepsilon)$ from the origin. Mathematical results on
%Helmholtz resonators can be found in \cite{gadilshin}. From
%Theorem \ref{apl} it follows that one can choose impedance $Z(t)$
%which depends continuously on $t$ and such that $Z(0)=0$
\begin{center}
\textbf{Proofs.}
\end{center}

We will start with a couple of general formulas needed below. First, let us recall the well known formulas for the scattering amplitude and total cross section:
   \begin{equation}\label{wknown}
  u_{\infty}(\theta)=\frac{1}{4\pi}\int_{|r|=R}\left  (ik(\theta \cdot \frac{r}{R}) u + u_r \right)e^{-ik(\theta \cdot r)} dS(r), \quad
  \sigma_k(u)=\frac{1}{k} \Im (u_n,u),
   \end{equation}
where $k>0$ and $R$ is large enough, so that the ball $|r|<R$ contains the obstacle.

The Green formula for the solutions of (\ref{1}) implies
\begin{equation}\label{Grf}
\int_{\Omega} \left [-|\nabla u|^2+k^2|u|^2 \right
]dx=\int_{\partial\Omega}u_n\overline{u}dS-ik\sigma_k(u),~k>0.
\end{equation}
Note that the second formula in (\ref{wknown}) is an obvious consequence of (\ref{Grf}).
Formula (\ref{Grf}) remains valid if $k=0$ and the second condition in (\ref{1}) is replaced by the decay of $u$ at infinity. In this case,
\begin{equation}\label{Grf0}
-\int_{\Omega}|\nabla u|^2dx=\int_{\partial\Omega}u_n\overline{u}dS,~k=0.
\end{equation}
Thus operator $\mathcal D(0)$ is negative.

\textbf{Proof of Theorem \ref{td}.} Operator $\mathcal D$ is compact in both spaces $L_2(\partial\Omega)$ and $H^{1/2}(\partial\Omega)$,
and condition $\frac{1}{\gamma(k)} \not \in Sp(\mathcal D(k))$ implies the solvability of (\ref{spectr}), and therefore, the existence of the
solution of problem (\ref{1})-(\ref{2}).

 Let $u$ be a solution of the non-homogeneous problem (\ref{1}) with a right-hand side of the equation in the space $L^2_{\rm{com}}(\Omega)$ (functions from $L^2$ with compact supports) and with zero Neumann boundary condition. Function $u$, being considered as element of the Sobolev space $H^2_{\rm{loc}}(\Omega)$, admits a meromorphic continuation to the whole complex plane $k \in \mathbb C $ with all the poles located in the lower half plane, see
\cite{v, Vai75}. Indeed, a pole at a point $k=k_0, \Im k_0>0,$ would lead to a complex eigenvalue of the Neumann Laplacian, a real pole $k=k_0 \neq 0$ would lead to  a non-uniqueness of the solution of problem (\ref{1}) with the Neumann boundary condition, and a pole at the origin would lead to the existence of a decaying at infinity harmonic function in $\Omega$ with zero Neumann boundary condition, see \cite{Vai75}. This properties of $u$ immediately imply that operator
\begin{equation}\label{anal}
\mathcal
D(k):H^{1/2}(\partial\Omega)\rightarrow H^{3/2}(\partial\Omega).
\end{equation}
is meromorphic in the whole $k$-plane with all the poles in the lower half plane. (Recall that operator $D(k)$ for complex k is understood as an analytic continuation from the operator on semiaxis $k>0$).

To obtain the same analytic properties of operator $D(k)$ in $L^2(\partial \Omega)$,
one needs to study the kernel $G_k$ of the operator (see
(\ref{dirGreen})). A standard approach to the construction of the
Green function (reduction to an integral equation on the boundary) allows one to show that $G_k=G_0+(G_k-G_0)$ where
$G_0$ is the Green function of the Neumann problem for the
Laplacian, function $G_k(x,s)-G_0(x,s)$ has a weak singularity at $x=s$, and the operator in $L^2(\partial \Omega)$ with the kernel $G_k(x,s)-G_0(x,s)$ is meromorphic in $k\in \mathbb C$. Poles in the region $\Im k \geq 0$ do not exist by the same reason as for operator (\ref{anal}).

The analytic Fredholm theorem implies that $(I-\gamma(k)\mathcal
D(k))^{-1}$ is meromorphic in $k\in U$ if the latter operator is bounded for some value of $k\in U$. In fact, one needs to refer to a Fredholm theorem for a meromorphic family of operators, see \cite {bl}.

Let us prove the second statement of the theorem. The kernel $G_k$ is real and symmetric when $k=i\rho, \rho \geq 0$. Thus the operator $\mathcal D(i\rho)$ is symmetric. If a family of compact operators (which is $\mathcal D(k)$ in our case) is symmetric on a ray and analytic in its neighborhood, then the eigenfunctions and eigenvalues of this family are analytic on the ray, see
\cite{K}. In particular, the eigenfunctions and eigenvalues of $\mathcal D(k)$ are analytic at $k=0$. Let us give a little more detail.

Formula (\ref{Grf0}) implies that $\varsigma_j(0)<0$. Besides, $\varsigma_j(0)\rightarrow 0$ as $j\rightarrow \infty $ since operator $\mathcal D(0)$ is compact.
Let us fix $\alpha \in (\varepsilon /2, \varepsilon)$ which is not an eigenvalue of $\mathcal D(0)$. Let $\Gamma_1$ be a bounded contour in the half plane Re$k<-\alpha$ which encloses all the eigenvalues  $\varsigma_j(0)<-\alpha$ and let $\Gamma_2$ be a circle of radius $\alpha$ centered at $k=0$ (which encloses the remaining eigenvalues). Obviously, operators
\[
P_{i}(k)=\frac{1}{4\pi i}\int_{\Gamma _i}(z-\mathcal D(k))^{-1}dz,~i=1,2,
\]
 are analytic in $k$ in a small neighborhood $U_{\varepsilon}$ of the origin, commute with $\mathcal D(k)$, and $P_{1}(k)+P_{2}(k)=I$ is the identity operator. We define $ L_{1,\varepsilon}(k)$ as the range of operator $P_{i}(k)$. Spaces $ L_{1,\varepsilon}(i\rho), \rho \geq 0,$ are spanned by the corresponding sets of eigenfunctions of the symmetric operator $\mathcal D(i\rho)$, and  $P_{i}(i\rho),~\rho \geq 0,$ are orthogonal projections, and therefore they are self-adjoint.  Thus eigenfunctions of $P_{1}(i\rho)$ (which are also  eigenfunctions of $\mathcal D(i\rho)$) admit an analytic continuation
 (see \cite{K}). We refer to the same source \cite{K} but in a simpler situation of a finite dimensional operator (one also could reduce the statement above to the statement for a matrix which is symmetric on a segment and analytic in a neighborhood of the segment). The norm of  $\mathcal D(k)P_{2}(k)$ does not exceed $\alpha <\varepsilon$ when $k=0$, and therefore it does not exceed $\varepsilon$
 when $k \in U_{\varepsilon}$ if $U_{\varepsilon}$ is small enough.
 The second statement of the theorem is proved.

 Let us prove the last statement. It is well known that the operator $\sigma : u|_{\partial\Omega}\rightarrow u_{\infty}$ is bounded (and compact) in $L^2$, i.e.,
 \begin{equation}\label {ot}
 \sigma_k(u) \leq C(k)||u||^2_{L^2(\partial\Omega)},~k>0.
 \end{equation}
Indeed, consider the operators which map Dirichlet data on
$\partial\Omega$ into the solution $u$ of the Dirichlet problem
and its derivative $u_r$ on the sphere $r=R, R\gg 1$. These
operators are given by formulas similar to (\ref{dirGreen}).
The kernels of these operators are smooth when $s \in
\partial\Omega, |r|=R$, and therefore the operators are bounded
(and compact). Thus an application of the first formula in (\ref{wknown}) implies
(\ref{ot}). On the other hand, from the second formula in (\ref{wknown}) it follows that
$\Im \varsigma_j^{-1}(k)||u_j||^2=k\sigma_k(u_j)$,  where $u_j$ is the
solution of the problem (\ref{1}) whose Neumann data is an
eigenfunction of $\mathcal D(k)$ with the eigenvalue
$\varsigma_j(k)$. This and (\ref{ot}) imply that $0<\Im \varsigma_j^{-1}<c(k)$ for some $c(k)<\infty$ and all $j$.
The estimate $\Re \varsigma_j^{-1}<d(k)$ follows from the representation
 \[
 D^{-1}(k)=D^{-1}(0)+[D^{-1}(k)-D^{-1}(0)],
 \]
where the first operator in the right-hand side is negative (see (\ref{Grf0})) and the second one is bounded (and compact). The compactness of the second operator can be easily derived from the fact that the kernel of operator $D^{-1}(k)$ is the normal derivative of the Green function of the Dirichlet problem. The theorem is
proved.

 \textbf{Proof of Theorem \ref{teorD}}.  Condition (\ref{2}) implies
 \begin{equation}\label{nd}
(\mathcal N(k)-te^{i\delta})u^t=f_1+te^{-i\delta}f,
~~~f=e^{ik(r\cdot \alpha)},~f_1=-\frac{\partial e^{ik(r\cdot
\alpha)}}{\partial n},
 \end{equation}
 where $\mathcal N(k)=\mathcal D^{-1}(k)$ is the unbounded operator in
 $L^2(\partial\Omega)$ which corresponds to the Dirichlet-to-Neumann
 map. Let us write it in the form $\mathcal N =A+iB$,
 where operators $A$ and $B$ are self-adjoint. Hence, we can rewrite (\ref{wknown}) as $(Bu,u)=\frac{1}{k}\sigma(u)$, and (\ref{ot}) implies that operator $B$ is bounded.

 We rewrite
 (\ref{nd}) in the form
 \begin{equation}\label{nd1}
u^t=(\mathcal N(k)-te^{i\delta})^{-1}(f_1+te^{i\delta}f)=
(I+i(A-te^{i\delta})^{-1}B)^{-1}(A-te^{i\delta})^{-1}(f_1+te^{i\delta}f).
 \end{equation}

Consider first the case of $\sin \delta \neq 0$. Since operator $A$ is self-adjoint, we have
\begin{equation}\label{aestim}
\|(A-te^{i\delta})^{-1}\|_{L_2(\partial \Omega)} \leq
\frac{1}{t|\sin(\delta)|}, \quad \sin (\delta) \neq 0.
\end{equation}
Thus
\begin{equation}\label{aestim2}
(I+i(A-te^{i\delta})^{-1}B)^{-1} \rightarrow I, \quad t
\rightarrow \infty, \quad \sin (\delta) \neq 0.
\end{equation}
From (\ref{aestim}) and the identity
\[
te^{i\delta}(A-te^{i\delta})^{-1}h=(A+te^{i\delta})^{-1}Ah-h
\]
which is valid for any $h$ in the domain of $A$, it also follows
that
\begin{equation}\label{ll}
s\!\!-\!\!\lim_{t \rightarrow \infty}
te^{i\delta}(A-te^{i\delta})^{-1} = I, \quad \sin (\delta) \neq 0.
\end{equation}
Since functions $f,f_1$ are smooth and belong to the domain of
$A$, relations (\ref{nd1}), (\ref{aestim}) and (\ref{ll}) imply
the first of relations (\ref{perehod}). The second relation
follows from the first one and (\ref{ot}).

Case of $\delta=0$ is treated similarly. One needs only to note that operator $A$ is bounded from above \cite[4.2]{agr} and therefore,
estimate (\ref{aestim}) holds with the right-hand side replaced by $1/(t-d), t>d$, where $d$ is a constant.

The proof is complete.

The following lemma will be needed for the proof of Theorem
\ref{t3}. Recall that operator $ \mathcal D(k)$ is analytic, i.e.,
\[ \mathcal D=\sum_{m=0}^{\infty} \mathcal D_m k^m.
\]
Obviously, the kernel of operator $\mathcal D_0$ is the Green
function of the Neumann problem for the Laplacian in the exterior
of $\Omega$. Let us show that
 $\mathcal D_1$ is the one dimensional operator of the projection
 on the constant
 $u_0$:
\begin{lemma}\label{green}
\[ \mathcal D_1v=-\frac{i}{4\pi}(v,u_0)u_0, \quad v \in
L_2(\partial \Omega,dS).
\]
\end{lemma}
\textbf{Proof.} It is enough to prove the statement above for smooth $v$ or $v \in H^{1/2}(\partial\Omega)$. Let $u$ be a solution of problem (\ref{1}) with the Neumann data at the boundary equal to $v$, i.e.,
\begin{equation}\label{aa}
(\Delta+k^2)u=0, ~x\in \Omega; ~~\frac{\partial u}{\partial n}|_{\partial\Omega}=v,
\end{equation}
We also assume that $u$ satisfies the radiation condition (\ref{1}). Green's formula implies that
\[
u(r_0)=\int_{\partial\Omega} \left [\frac{\partial}{\partial
n}(\frac{e^{ik|r-r_0|}}{4\pi |r-r_0|})u-\frac{e^{ik|r-r_0|}}{4\pi
|r-r_0|}v \right ]dS,~~r_0 \in \Omega.
\]

If $v \in H^{1/2}(\partial\Omega)$, then solution $u=\sum_0^{\infty} k^ju_j$ is analytic in $k$ as an element of $ H^{2}(\Omega)$, see  \cite{v}. We expand the left and right hand sides in the formula above in the Taylor series and equate the linear terms. Then we arrive at
\begin{equation}\label{ab}
u_1(r_0)=\int_{\partial\Omega}\left [\frac{\partial}{\partial
n}(\frac{1}{4\pi |r-r_0|})u_0-\frac{i}{4\pi }v \right ]dS,~~r_0
\in \Omega.
\end{equation}
From this equation it follows that $|u_1|<C<\infty$ as $r_0 \rightarrow\infty$. On the other hand, (\ref{aa})
implies that $\triangle u_1=0$ in $\Omega$, $(u_1)_n=0$ on $\partial\Omega$. Thus, $u_1$ is constant in $\Omega$. The integral of the first term in the right hand side of (\ref{ab}) decays at infinity, and the integral of the send term is a constant. Hence the first integral is zero, and
\[
u_1=-\frac{i}{4\pi }\int_{\partial\Omega}vdS,~~r_0 \in \Omega.
\]
It remains to note that $\mathcal D_1v=u_1|_{\partial\Omega}$. The proof is complete.

\textbf{Proof of Theorem \ref{t3}.} The integral kernel of
operator $-\mathcal D(0)$ is positive (see, e.g., \cite[Example 3,
XIII.12]{rs}). Thus, from the Perron-Frobenius theorem it follows that
$\varsigma_0$ is simple and the sign of the corresponding
eigenfunction is not changing. The latter implies that $c_0 \neq
0$. Let us prove (\ref{cp}). Let $u$ be a harmonic function in
$\Omega$ equal to $u_0\equiv 1$ on $\partial\Omega$. Then
\[
C=-\frac{1}{4\pi}\int_{\partial\Omega}\frac{\partial u}{\partial
n}dS=\frac{1}{4\pi}\left (-\mathcal Nu_0,u_0 \right ),~~ \mathcal
N=\mathcal D^{-1}(0).
\]
Thus,
\[
\frac{-1}{\varsigma_0(0)}=\min_{\varphi\neq 0}\frac{(-\mathcal N\varphi,\varphi)}{||\varphi||^2}\leq
\frac{(-\mathcal Nu_0,u_0)}{||u_0||^2}=\frac{4\pi C}{S}.
\]
The first statement is proved. Let us prove the second
statement.

Since operator $\mathcal D (k)$ is analytic and
$\mathcal D (0)$ is symmetric, the standard perturbation theory
implies that
\[
\varsigma'_p(0)=(\mathcal D'(0) \varphi_p(0),\varphi_p (0))=(\mathcal D_1 \varphi_p(0),\varphi_p (0)),
\]
and Lemma \ref{green} justifies statement 2.

Let us prove the last two statements. Since operators (\ref{anal}) are bounded and analytic in $k$, operators
\[
\mathcal D(k):H^{1/2}(\partial\Omega)\rightarrow
H^{3/2}(\partial\Omega)
\]
are compact and analytic in $k$. Thus, if $\frac{1}{\gamma(0)}
\not \in Sp(\mathcal D(0))$, then
the analytic Fredholm theorem implies that the solution $v \in H^{1/2}(\partial\Omega)$
of equation (\ref{spectr}) exists and is analytic in $k$ in some neighborhood of $k=0$, and therefore (see \cite{v})
solution $u \in H^2_{loc}(\Omega)$ of (\ref{1})-(\ref{2}) exists in some neighborhood of
$k=0$ and is analytic in $k$.

Let now $\frac{1}{\gamma(0)}=\varsigma_p(0)$ and (\ref{gpr}) hold. We fix $\varepsilon=-\gamma(0)/2$, split $L_2(\partial \Omega)$ according to (\ref{split}) and
represent the right hand side $f$ in equation (\ref{spectr}) in the form
\[
f=f_1+f_2,~~ f_1=P_1(k)f \in L_{1,\varepsilon},~f_2=P_2(k)f \in L_{2,\varepsilon}.
\]
Then solution $v\in H^{1/2}(\partial\Omega)$ has the form
\[
v=v_1+v_2,~~v_i=(I-\gamma(k)\mathcal D(k))^{-1}_iP_i(k)f,
\]
where subindex $i$ in the first operator on the right indicates the restriction of the operator on the space
$L_{i,\varepsilon}(k)$:
\[
(I-\gamma(k)\mathcal D(k))^{-1}_i=
(I-\gamma(k)\mathcal D(k))^{-1}:L_{i,\varepsilon}
\rightarrow L_{i,\varepsilon}.
\]

From the choice of $\varepsilon$ it follows that the operator
\[
(I-\gamma(k)\mathcal D(k))^{-1}_2=\sum_{j=0}^{\infty}(\gamma(k)\mathcal D(k))^j
\]
is analytic in a neighborhood of $k=0$, and therefore the same is true for $v_2\in H^{1/2}(\partial\Omega)$.
In order to find $v_1$, we write $P_1(k)f$ in the form
\[
P_1(k)f=\sum_{j=0}^{N_{\varepsilon}}a_j(k)\varphi_j(k),~~a_j(0)=\gamma(0)c_j.
\]
Then, as $k \rightarrow 0$, we have
\begin{equation}\label{vvv}
v_1(r)=\sum_{j=0}^{N_{\varepsilon}}\frac{1}{1-\gamma(k)\mathcal D(k)}a_j(k)\varphi_j(k,r)=
\sum_{j=0}^{N_{\varepsilon}}\frac{1}{1-\gamma(k)\mathcal D(k)}(\gamma(0)c_j\varphi_j(0,r)+O(k)).
\end{equation}
Hence,
\begin{equation}\label{vv}
v=\frac{1}{k}\sum _p^{p+m-1}\frac{-c_j \gamma(0) }{(
\gamma\varsigma_j)'(0)}\varphi_j(0,r)+ O(1),~~k\rightarrow 0.
\end{equation}
Since $u=\mathcal Dv,~ r \in \partial\Omega$, (\ref{vv}) immediately implies (\ref{ass}).

In order to obtain (\ref{assSigma}), we note that the Green formula allows us to rewrite
the first equality in (\ref{wknown}) in the form
\[
  u_{\infty}(\theta)=\frac{1}{4\pi}\int_{\partial\Omega} \left (u\frac{\partial}{\partial n}
  e^{-ik(\theta \cdot r)}- ve^{-ik(\theta \cdot r)} \right ) dS. \quad
\]
This, (\ref{vv}) and (\ref{ass}) imply that, as $k\rightarrow 0$,
\[
u_{\infty}(\theta)=\frac{-1}{4\pi}\int_{\partial\Omega}vdS+O(1)=
\frac{-1}{4\pi k}\sum _p^{p+m-1}\frac{-c_j^2 \gamma(0) }{(
\gamma\varsigma_j)'(0)}+ O(1) =\frac{\gamma(0)}{4\pi k}\sum
_p^{p+m-1}c_jb_j  + O(1),
\]
where coefficients $b_j$ are defined in (\ref{bjDef}). This proves
(\ref{assSigma}). The last statement of the theorem follows from
analyticity of $v_2$, (\ref{vvv}) and linear independence of
functions $\varphi_j$.

The proof is complete.

\textbf{Proof of theorem \ref{tlast}}. The first statement is an analytic in $\gamma$ extension of (\ref{unitar}) ($\gamma_1$ appears in that statement because the integral kernel of operator $F^*$ contains complex conjugation). Let us prove the second statement. From (\ref{wknown}) it follows that the integral kernel
$u_{\infty}$ of operator $F$ is smooth with respect to $\theta, \alpha$ and meromorphic in $k$, i.e.,
(\ref{scat}) is a meromorphic family of Fredholm operators. Consider $\mathcal S_{\gamma}^{-1}(k)$. From the relation
$\mathcal S_{\gamma}^{-1}(k)\mathcal S_{\gamma}(k)=I$ it follows that $\mathcal S_{\gamma}(k)$ has a pole at $k=k_0$ if and only if $\mathcal S_{\gamma}^{-1}(k)$ has a non-trivial kernel at this point.

Consider now operator $\widehat{\mathcal S}_{\gamma}(k)$ which is
defined as follows. If $\mathcal
S_{\gamma}^{-1}(k)f(\theta)=g(\theta)$, then $\widehat{\mathcal
S}_{\gamma}(k)f(-\theta)=g(-\theta)$. Obviously, operators
$\mathcal S_{\gamma}^{-1}(k)$ and $\widehat{\mathcal
S}_{\gamma}(k)$ have non-trivial kernels at the same points $k \in
\mathbb C$.

Let us construct operator $\widehat{\mathcal S}_{\gamma}(k)$. First we assume that $k>0$. From (\ref{sscc}) it follows that
\[
\mathcal S_{\gamma}^{-1}(k)u_{out}(\theta)=u_{in}(-\theta),~~k > 0.
\]
Thus
\[
\widehat{\mathcal S}_{\gamma}(k)u_{out}(-\theta)=u_{in}(\theta),~~k > 0.
\]
On the other hand, after complex conjugation in (\ref{1234}) and application of (\ref{sscc}), we get
\[
\mathcal S_{\gamma_1}(k)\overline{u_{out}}(-\theta)=\overline{u_{in}}(\theta),~~\gamma_1=\overline{\gamma},~~k > 0.
\]
Hence,
\[
\widehat{\mathcal S}_{\gamma}(k)f(\theta)=\overline{\mathcal S_{\gamma_1}(k)\overline{f}(\theta)},~~\gamma_1=\overline{\gamma},~~k>0.
\]
We extend the last relation analytically in the complex plane $k$.
If $\gamma_1=\overline{\gamma(\overline{k})}, ~~k\in \mathbb C,$
we obtain that
\[
\widehat{\mathcal S}_{\gamma}(k)f(\theta)=\overline{\mathcal
S_{\gamma_1}(\overline{k})\overline{f}(\theta)},~~k \in \mathbb C.
\]
This proves the second statement of the theorem. The last statement follows immediately from Theorem \ref{td}. The proof is complete.

\begin{center}
 \textbf{Appendix 1. }
\end{center}
\textbf{A. Reduction of acoustic equations to an impedance problem.} The general acoustic equations in $\Omega=R^3\backslash\mathcal O$ have the form
\begin{equation}\label{acou}
\Delta p(r)+k^2p(r)=0,~~ -i\omega\rho_l(\nabla u)=-\nabla p,
\end{equation}
where $p$ satisfies the radiation conditions at infinity. Here
$Re(p(r)e^{-i\omega t})$ is the pressure, $u$ is the velocity
potential ($v=\nabla u$), $\rho_l>0$ is the density of the media.
The second equation above follows from the Newton law: density of
the media times acceleration equals to the negative of the
gradient of the pressure, where the minus sign is needed since the
direction of the force corresponds to the decay of the pressure.

Now we are going to derive boundary conditions for equations (\ref{acou}) assuming that the obstacle is covered by a springy layer with mutually independent "springs", and the thickness $h\sim 0$ of the layer is negligible. Since the normal velocity of the media at the boundary coincides with the normal velocity of the cover of the obstacle (there are no voids between media and the cover), the Hooke's law implies
\begin{equation}\label{25}
\frac{1}{-i\omega}\frac{\partial u}{\partial n}(r)=-\beta p(r),~ r\in \partial \Omega.
\end{equation}
Here $n$ is the interior unit normal vector to
$\partial\Omega$, the left hand side is the radial
displacement (the integral with respect to time of the normal
velocity), $\beta >0$ is the compressibility coefficient (for
"springs"), and the sign minus on the right indicates that the
layer shrinks when $\beta>0$ increases. Note that $\beta$ could
depend on the point $r\in \partial\Omega$.

The second equation in (\ref{acou}) implies that
\begin{equation}\label{26}
\frac{\partial u}{\partial n}=\frac{1}{i\omega \rho_l}\frac{\partial p}{\partial n}.
\end{equation}
From here and the Hooke law it follows that
\begin{equation}\label{zz}
\frac{\partial p}{\partial n}(r)+\beta\rho_l\omega^2p(r)=0,~~r\in\partial\Omega.
\end{equation}
We obtained the impedance boundary condition for $p$, where the impedance has the form (\ref{imz}) with $Z=\beta\rho_l$.

\textbf{B. Evaluation of the compressibility coefficient.} Consider a rigid obstacle $\mathcal O$ covered by an elastic membrane with a gas (for example, air) layer in between them and numerous rigid walls fixed to the obstacle and the membrane which partition the gas layer into small chambers. We assume that the sizes of chambers are small enough so that the impedance can be considered as local. All the construction is submerged into a liquid. We will show that the compressibility coefficient $\beta$ is given by the following expression
\begin{equation}\label{l1}
\beta=\frac{\gamma_g}{\rho_g c_g^2}h,
\end{equation}
where $\rho_g$ the density of the gas in the chambers,
$c_g$ is the speed of the sound propagation there,
$h$ is the distance between the obstacle and the membrane,
$\gamma_g=C_p/C_v$ is the ratio of heat capacities of the gas with constant
pressure and volume respectively.

In fact, Kliperon's law implies that
\begin{equation}\label{l2}
P V = \frac{m}{\mu} RT,
\end{equation}
where $P,V$ are values of unperturbed pressure and volume in the chambers,
$\mu$ is molecular weight of the gas, $R$ is the universal gas constant. Let $\Delta P, \Delta V$ be the change in $P$ and $V$ respectively.
Assuming that the temperature is constant, we get
\begin{equation*}
\frac{\Delta V}{\Delta P} = -\frac{V}{P}.
\end{equation*}
Let $V=Sh, \quad \Delta V  = S \Delta h$, where $S$ is the area of the chamber base. Then
\begin{equation}\label{l4}
\beta=-\frac{\Delta h}{\Delta P} = \frac{h}{P},
\end{equation}
From (\ref{l2}) it follows that
\begin{equation*}
P=\frac{m}{V} \frac{RT}{\mu} = \rho_g \frac{RT}{\mu},
\end{equation*}
and since
\begin{equation*}
c_g = \sqrt{\frac{\gamma_g RT}{\mu}},
\end{equation*}
we obtain
\begin{equation}\label{l7}
P=\frac{\rho_g (c_g)^2}{\gamma_g}.
\end{equation}
We substitute the latter expression into (\ref{l4}) and arrive at
(\ref{l1}).

From (\ref{l1}) and (\ref{zz}), taking into account that $\omega=kc_l$, we obtain the following final
expression for the impedance in the problem under consideration
\begin{equation}\label{impgl2}
\gamma=-\gamma_g\frac{\rho_l}{\rho_g} \left ( \frac{c_l}{c_g}
\right )^2 h k^2.
\end{equation}

Let us find this value assuming that the air is used as a gas and water with zero temperature is outside of the construction. Then
$$
\begin{array}{ccccc}
\hline  \gamma_g  & \rho_g   & c_g    & \rho_l   & c_l   \\
\hline
 1.383 & 1.3 kg/m^3 & 331.3 m/s & 1000 kg/m^3 & 1390 m/s \\ \hline
 \end{array}
$$
and therefore
\begin{equation}\label{l8}
\gamma \sim - 25800 h k^2.
\end{equation}
 Note that both $k$ and $\gamma$ are measured in unites of (length)$^{-1}$.

\textbf{C. Impedance when a friction is present.} Assume that oscillations of
 a springy cover is accompanied by friction. Usually,
 a force due to friction is proportional to the velocity: $F_{fr}=-\varepsilon du/dn$.
In natural circumstances, $\varepsilon \geq 0$, but one also can create artificially a
 situation when $\varepsilon < 0$ (negative friction). Let us write an analog of (\ref{25}).
  By equating all the forces, we obtain
\begin{equation}\label{6}
-\varepsilon \frac{\partial u}{\partial n} +
\frac{-1}{\beta}\frac{1}{-i \omega}\frac{du}{dn}(r)= p(r), \quad r
\in
\partial \Omega.
\end{equation}
Using (\ref{26}), we obtain
\begin{equation}\label{7}
\frac{dp}{dn}(r)+ (\beta \rho_l) \omega^2 \left (
\frac{1+i\varepsilon \omega \beta}{1+(\varepsilon \omega
\beta)^2}\right ) p(r) =0 , \quad r \in
\partial \Omega.
\end{equation}

One can see that the presence of a friction creates an imaginary part of $\gamma$, and natural friction corresponds to
$\Im \gamma <0$, and "negative friction" corresponds to $\Im \gamma >0$.

\bigskip

\noindent \textbf{Acknowledgement.} The authors are very grateful
to N. Grinberg, A. Kirsch and M.A. Mironov for numerous and useful
discussions. In particular, appendix I would not have been written without
an input from M.A. Mironov.

The work of the first author was
supported in part by {\it Centre for Research on Optimization and
Control} (CEOC) from the ''{\it Funda\c{c}\~{a}o para a
Ci\^{e}ncia e a Tecnologia}'' (FCT), cofinanced by the European
Community Fund FEDER/POCTI, and by the FCT research project
PTDC/MAT/113470/2009. The work of the second author was supported
in part by the NSF grant DMS-0706928.


\begin{thebibliography}{99}
\bibitem{arx} E. Lakshtanov, Spectral properties of the Dirichlet-to-Neumann operator for exterior Helmholtz problem and its applications to scattering
theory, Submitted.
\end{thebibliography}
\end{document}